\documentclass[10pt, conference,compsocconf]{IEEEtran}  

\usepackage[bottom]{footmisc}
\usepackage{ifpdf}
\usepackage{cite}
\ifCLASSINFOpdf        
   \usepackage[pdftex]{graphicx}
\else
   \usepackage[dvips]{graphicx}
\fi
\usepackage[cmex10]{amsmath}   
\usepackage{algorithmic}
\usepackage{array}
\usepackage{bm}
\usepackage{mdwmath}
\usepackage{mdwtab}
\usepackage{eqparbox}
\usepackage{url}
\usepackage{subfigure}
\usepackage{balance}
\usepackage{multirow}
\usepackage{array}       
\usepackage[numbers,square,comma,sort&compress]{natbib}   
\usepackage[lined, algonl, boxed]{algorithm2e}
\usepackage{booktabs}
\usepackage{float}
\usepackage{dashrule}

\usepackage[final,inline,nomargin]{fixme}      
\usepackage{color}
\usepackage{soul}          

\newenvironment{smitemize}
  {\begin{list}{$\bullet$}
     {\setlength{\parsep}{0pt}
      \setlength{\leftmargin}{10pt}
      \setlength{\topsep}{-7pt}
      \setlength{\labelwidth}{5pt}
      \setlength{\itemsep}{1pt}}}
  {\end{list}}

\begin{document}

\title{A Comprehensive Study of the GeoPass User Authentication Scheme}

\author{
      \IEEEauthorblockN{Mahdi Nasrullah Al-Ameen and Matthew Wright\\}
      \IEEEauthorblockA{Department of Computer Science and Engineering\\
                 The University of Texas at Arlington \\
                 Arlington, TX, USA\\
              mahdi.al-ameen@mavs.uta.edu, mwright@cse.uta.edu}}

\maketitle


\begin{abstract}
Before deploying a new user authentication scheme, it is critical to
subject the scheme to comprehensive study. Few works, however, have
undertaken such a study. Recently, Thorpe et al. proposed GeoPass, the
most promising of a class of user authentication schemes based on
geographic locations in online maps. Their study showed very high
memorability (97\%) and satisfactory resilience against online guessing,
which means that GeoPass has compelling features for real-world use. No
comprehensive study, however, has been conducted for GeoPass or any
other location-based password scheme. In this paper, we present a
systematic approach for the detailed evaluation of a password system,
which we implement to study GeoPass. We conducted three separate studies
to evaluate the suitability of GeoPass for widespread use. First, we
performed a field study over two months, in which users in a real-world
setting remembered their location-passwords $96$\% of the time and
showed improvement with more login sessions. Second, we conducted a
study to test how users would fare with multiple location-passwords and
found that users remembered their location-passwords in less than $70\%$
of login sessions, with $40\%$ of login failures due to interference
effects. Third, we conducted a study to examine the resilience of
GeoPass against shoulder surfing. Our participants played the role of
attackers and had an overall success rate of $48$\%. Based on our
results, we suggest suitable applications of GeoPass in its current
state and identify aspects of GeoPass that must be improved before
widespread deployment could be considered.

\end{abstract}




\section{Introduction} 
Users often choose passwords that are easy to remember but also easy to guess. Both textual~\cite{passphrase,text_recog,forget_thesis} and graphical passwords~\cite{das,passpoint1,story,passface,survey} have failed to provide a viable solution to this usability-security tension. It thus remains a critical challenge in password research to design an authentication scheme that is resilient to guessing attacks while offering good memorability.

Thorpe et al.~\cite{geopass} presented a possible breakthrough with GeoPass in 2013. They show that this geographic location-password scheme offers resilience to online guessing attacks while providing very good memorability ($97$\%, found in a nine-day-long lab study) and high user satisfaction. While GeoPass shows potential to possibly solve the usability-security tension in user authentication, comprehensive study is required before widespread deployment. In this paper, we present the design and results of such a comprehensive study aimed at investigating the suitability of GeoPass for real-world applications.

\subsection{Motivations}

After decades of studies on passwords, we are still looking for an authentication scheme that would more effectively solve the usability-security tension in user authentication. Biddle et al.~\cite{survey} performed a comprehensive survey on $25$ different graphical password schemes and found that the research community still remains unsure of the applicability of proposed schemes in the real world due to the lack of comprehensive studies conducted in an organized way. Their analysis and findings inspired us to carry out such a comprehensive study to examine the real-world applicability of a scheme that, while promising, had only been tested in a single lab study.

In this paper, we first spell out the research questions for the comprehensive analysis of an authentication scheme. We then design and implement three independent studies addressing those research questions in an organized way. We are motivated by the following facts to evaluate GeoPass through our study design: i) GeoPass provides a potential avenue to solving the usability-security tension in user authentication~\cite{geopass}, and ii) The lessons from first twelve years of graphical passwords~\cite{survey} make a point clear that unless the primary usability and security issues of a new category of passwords are identified at the initial phase, the later schemes in that category might have novelty in certain aspects but would not be able to effectively address the usability-security tension in user authentication. So, it is important to better understand the design space for geographical password schemes, the most recent inclusion in password research. 
 
\subsection{Contributions}

We identify the following research questions for GeoPass to examine its potential for deployment:
\vspace{0.2cm}
\begin{smitemize}
\item How usable would GeoPass be in a real-world setting?
\item How do the login performances in GeoPass change over login sessions (i.e.,
  what is the training effect)?
\item How usable would GeoPass be when users would have to remember
  multiple location-passwords?
\item How prominent will the interference effects be for multiple
  location-passwords?
\item How resilient is GeoPass against shoulder-surfing attacks?
\end{smitemize} 
\vspace{0.4cm}

We designed a systematic approach for addressing these research questions using three studies: a field study to address the first two questions, an interference study for the next two questions, and a shoulder-surfing study for the last question. 

We now present a high level overview of our findings with references to the corresponding sections that accommodate detailed discussions. In this paper, \textit{number of attempts} and \textit{login time}, respectively refer to the required attempts and time for successful logins only, unless otherwise specified.
 
\subsubsection{Field study (\S\ref{field})} 

Biddle et al.~\cite{survey} note that while field studies are challenging to perform, due to requiring a significant investment in resources and time, they offer strong ecological validity and the best measure of login performance in a realistic setting. We conducted a $66$-day-long field study including $1781$ login sessions from $50$ participants.

\textbf{Login performance.} GeoPass offered satisfactory memorability in our real-world context, with an overall login success rate of $96.1$\% and a median login time of $19$ seconds. We analyzed the login performance distribution among users for a detailed understanding of the usability issues (see~\S\ref{login_field}).

\textbf{Training effect.} A training effect occurs when users get better at entering their password over time. Prior field studies on graphical passwords lack a detailed analysis of training effect, even though it is of particular interest to the research community to learn how the login performances change during a long-term field study. Through probing the change in login performance over login sessions we demonstrate a training effect -- an overall improvement in login performance with more login sessions (see~\S\ref{training_field}). 

\subsubsection{Interference study (\S\ref{interference})} 

Biddle et al.~\cite{survey} identify multiple password interference as a major usability concern and make a surprising discovery that only three out of $25$ graphical password schemes have been evaluated with an interference study. While Biddle et al.~\cite{survey} highlight the importance of interference studies, they also report that how to best evaluate multiple password interference still remains an open issue. In this case, prior interference studies on graphical passwords showed the memorability for multiple passwords, but they failed to demonstrate the precise impact of interference (see~\S\ref{disc} for further discussion). 

We performed a three-week-long interference study in a lab setting, where the participants had to remember four different location-passwords, one for each of four different accounts. 

\textbf{Interference effects.} We identified that interference effects played a major role in the failure of login attempts when users had to remember multiple location-passwords. In this case, the overall login success rates were $58$\% in {\em login 1} (one week after registration) and $67$\% in {\em login 2} (two weeks after registration) (see~\S\ref{login_interference}).

Our detailed examination on login failures reveals the precise impact of interference. $44.8$\% of attempts in {\em login 1} and $38.2$\% of attempts in {\em login 2} failed because of interference effects (see~\S\ref{interference_effect}). We investigated both \textit{accurate} and \textit{non-accurate} interferences for an in-depth analysis of interference effects. 

\textbf{Correlation with distance.} Some participants created multiple geopasses around a particular area, perhaps to aid memorability. While a large fraction of login failures occurred because of interference effects, we investigated whether participants were confused by geopasses that were geographically close. Our analysis reveals that the interference effect between a pair of geopasses had no correlation with the distance between them (see~\S\ref{distance_interference}).

\textbf{Geopass creation behavior.} Through an anonymous paper-based survey, we identified the types of locations (e.g., home, workplace, etc.) participants chose as their location-passwords. We measured the distances between each given participant's geopasses to find the likelihood of creating geographically close location-passwords. We also discuss, based on our findings, possible geopass creation strategies for high security accounts, such as online bank accounts (see~\S\ref{spec}).

\subsubsection{Shoulder-surfing study (\S\ref{shoulder_surfing})} 

To examine the vulnerability of GeoPass to shoulder-surfing attacks, we conducted a lab study with $30$ participants playing the role of attacker.

\textbf{Success rate.} The overall success rate in shoulder surfing was $48$\%, which suggests that GeoPass is highly vulnerable to the attack (see~\S\ref{shoulder_login}). We analyzed the login attempts of unsuccessful participants. Our findings suggest that the success rate could further increase if they were allowed more login attempts, since the clicked locations of the participants who failed to log in successfully were very close to the actual location-password in several cases (see~\S\ref{shoulder_distance}).

\textbf{Recording materials.} Using the Google map interface on a tablet PC represents a tech-savvy method to record authentication secret during shoulder surfing, while pen and paper is traditionaly used in shoulder surfing studies. Participants in our study preferred pen and paper to using a tablet (see~\S\ref{shoulder_feedback}). 

\textbf{Navigation strategies.} We analyzed the impacts of navigation strategies on shoulder surfing and figured out that how a user navigated to the location-password, either by panning or typing the full address, did not affect the success rate of the attack (see~\S\ref{shoulder_login}).

\subsubsection{Overall recommendation}
Based on the results of our three studies and detailed analysis, we believe that GeoPass as proposed by Thorpe et al.~\cite{geopass} is not yet mature enough for wide-scale deployment. In particular, we examine the scenarios in which it may be useful and identify the aspects for future improvement (see~\S\ref{app}).

\section{Background}\label{background} 

In this section, we give a brief overview of notable textual and graphical password schemes, in which we highlight why existing schemes are insufficient. We then discuss geographic location-passwords and their potential. 

\subsection{Textual Password Schemes}
\vspace{0.1cm}
\textbf{Traditional passwords.} Traditional user-chosen textual passwords are fraught with security problems~\cite{yan_sec,interference4} and are especially prone to password reuse and predictable patterns~\cite{captcha,guessing08}. Different password restriction policies have been deployed to get users to create stronger passwords. Shay et al.~\cite{pwpattern1} report, however, that such policies do not necessarily lead to more secure passwords but do adversely affect memorability in some cases.

\textbf{Mnemonic passwords.} Kuo et al.~\cite{mnemonic} studied user-chosen passwords based on mnemonic phrases, which are slightly more resistant to brute-force attacks than traditional passwords. However, mnemonic passwords are predictable when users create passwords from common phrases. The vulnerability to guessing is further increased when attackers use a properly chosen dictionary~\cite{mnemonic}.

\textbf{System-assigned password.} System-assigned random textual password schemes are more secure but fail to provide sufficient memorability, even when natural-language words are used~\cite{passphrase,text_recog}.  Wright et al.~\cite{text_recog} compared the usability of three different system-assigned textual password schemes: Word Recall, Word Recognition, and Letter Recall. None of these had sufficient memorability rates. 

\textbf{PTP.} Persuasive Text Password (PTP)~\cite{persuation,forget_thesis} is a hybrid between user-selected and system-assigned passwords, in which the user first creates a password and PTP improves its security by placing randomly-chosen characters at random positions in the password. Unfortunately, the memorability for PTP is just $25$\% when two random characters are inserted~\cite{forget_thesis}.

\subsection{Graphical Password Schemes}

Graphical password schemes can be divided into three categories~\cite{survey}, based on the kind of memory leveraged by the systems: i) Drawmetric (recall-based), ii) Locimetric (cued-recall-based), and iii) Cognometric (recognition-based).

\textbf{Drawmetric.} The user is asked to reproduce a drawing in this category of graphical passwords. In \textit{Draw-a-Secret (DAS)}~\cite{das}, a user draws on top of a grid, and the password is represented as the sequence of grid squares. Nali and Thorpe~\cite{das2} have shown that users choose predictable patterns in DAS. \textit{BDAS}~\cite{bdas} intends to reduce the amount of symmetry in the user's drawing by adding background images, which may introduce other predictable behaviors such as targeting similar areas of the images or image-specific patterns~\cite{survey}. DAS and BDAS have recall rates of no higher than $80\%$.

\textbf{Locimetric.} The password schemes in this category, including {\em Passpoints} and {\em Cued Click-Points (CCP)}, present users with an image and have users select points on the image as their password. Dirik et al.~\cite{passpoint4} developed a model that can predict $70-80$\% of the user's click positions in Passpoints. To address this issue, Chiasson et al. proposed 
 \textit{Persuasive Cued Click-Points (PCCP)}~\cite{pccp,pccp2}, in which a randomly-positioned viewport is shown on top of the image during password creation, and users select their click-point within this viewport. The memorability for PCCP was found to be $83-94$\%. In a follow-up study, Chiasson et al.~\cite{click_pattern} found predictability in users' click points and indicate that the hotspot issue is still a security concern for PCCP.

\textbf{Cognometric.} In this recognition-based category of graphical passwords, the user is asked to recognize and identify their password images from a set of distractor images. \textit{Passfaces}~\cite{passface,passface2} is a commercial cognometric system in which users select one face among a panel of nine distractor faces and repeat this over several panels. Davis et al.~\cite{story} have found that users select predictable passwords on Faces (their own version of Passfaces), biased by race, gender and attractiveness of faces. As a result, the commercial Passfaces~\cite{passface} product now assigns a random set of faces instead of allowing users to choose. However, Everitt et al.~\cite{interference3} show that users have difficulty in remembering system-assigned Passfaces.

{\em Thus, despite a large body of research, it remains a critical challenge to build an authentication system that offers both high memorability and guessing resilience.}

\subsection{Geographic Location-Password} 

In 2013, Thorpe et al.~\cite{geopass} presented their study of a promising scheme for user authentication called GeoPass, which is based on selecting a location in an online map. They showed that the scheme offers high memorability and resistance against online guessing attacks, which suggests that GeoPass may offer a solution to the usability-security tension in password schemes.

GeoPass has the interesting property that it involves elements of recognition, cued-recall, and pure recall~\cite{geopass}, while a graphical password is generally categorized either as a recognition-based, recall-based or a cued-recall based scheme~\cite{survey}. 

In GeoPass, the user's password is a location on a digital online map (Google Maps). This secret location, known both as the {\em location-password} and just {\em geopass}\footnote{in lowercase to avoid confusion with the system name}, is selected by the user at registration by right-clicking on the map. GeoPass leverages the map display, search, zoom, panning, and marker placement features provided by the Google Maps API. 

The search bar helps to make navigation faster by enabling the user to type the name of a place and by providing a drop-down menu suggesting the locations in which the searched item may appear. Zooming and panning are also enabled via the Google Maps API. Using the convention that a higher numbered zoom level represents being zoomed in closer, the initial zoom level is 2, and GeoPass allows the user to click on a location at a minimum zoom level of $16$. A successful login requires the users to click within a $21$x$21$ pixel box around the location-password they had set. We refer readers to Thorpe et al.'s paper~\cite{geopass} for in-depth discussion on the features of GeoPass.

Thorpe et al.~\cite{geopass} conducted a nine-day-long user study on GeoPass with three sessions: two in a lab-setting and one online. The login success rate was $97$\%, and the median login time was found to be
no greater than $30$ seconds~\cite{geopass}. Their security analysis~\cite{geopass} showed that the theoretical password space for GeoPass is $2^{36.9}$, such that only $11$\% of online guessing attacks might be successful after allowing for $2^{16}$ guesses. In this respect, reasonable lockout rules~\cite{strongpw} should make GeoPass sufficiently resilient against such attacks. The user feedback on GeoPass was encouraging, and a number of users showed interest to use the scheme in real life~\cite{geopass}.

\textbf{Other schemes.} There are two other schemes that use map locations as an authentication secret: one proposed by Spitzer~\cite{geomap} and another one called PassMap~\cite{passmap}. PassMap requires the user to choose two locations and the scheme by Spitzer~\cite{geomap} requires five or seven locations at different zoom levels to be selected as the location-password. Thorpe et al.~\cite{geopass} have shown that GeoPass is more usable than other digital-map-based schemes~\cite{passmap,geomap} because of its requirement to click on a single location and normalized error tolerance to a given zoom level. The login success rate in GeoPass ($97$\%) was found to be higher than that in PassMap ($92.59$\%).

\section{Significance Tests}\label{test}
To analyze our results, we use statistical tests and consider results comparing two conditions to be significantly different when we find $p<0.05$. We selected statistical tests based on their appropriateness for the corresponding datasets. In this brief section, we give an overview of the statistical tests that we use in our analysis.

When comparing two conditions where the variable is at least ordinal, we use a Wilcoxon signed-rank test for the matched pairs of subjects and a Wilcoxon-Mann-Whitney test for unpaired results. Wilcoxon tests are similar to t-tests, but make no assumption about the distributions of the compared samples, which is appropriate to the datasets in our conditions. Whether or not a participant successfully authenticated is a binary measure. So, we use either a McNemar's test (for matched pairs of subjects) or a chi-squared test (for unpaired results) to compare login success rates between two conditions.

\section{Field Study}\label{field}
In this section, we describe the procedure and results of the field study,
which reveals the login performance of users in a real-world setting and
the changes in login performance over the login sessions.

\begin{figure}[t]
\centering
\includegraphics[width=85mm]{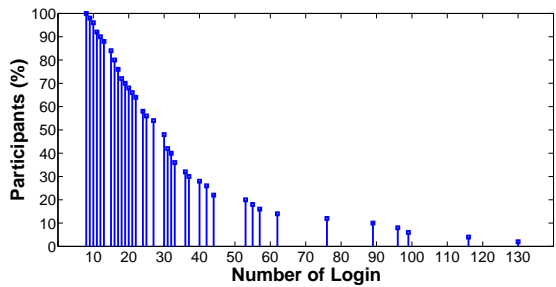}
\vskip -5 pt
\caption{{\em Field Study:} Number of logins}
\label{fig:num_login}
\vskip -4 pt
\end{figure}

\begin{figure*}[!t]
\vskip -10 pt
\centering
\subfigure[Success rate]{
\label{fig:success}
\includegraphics[width=0.32\linewidth,height=35mm,trim=1.5cm 0 4cm 0]{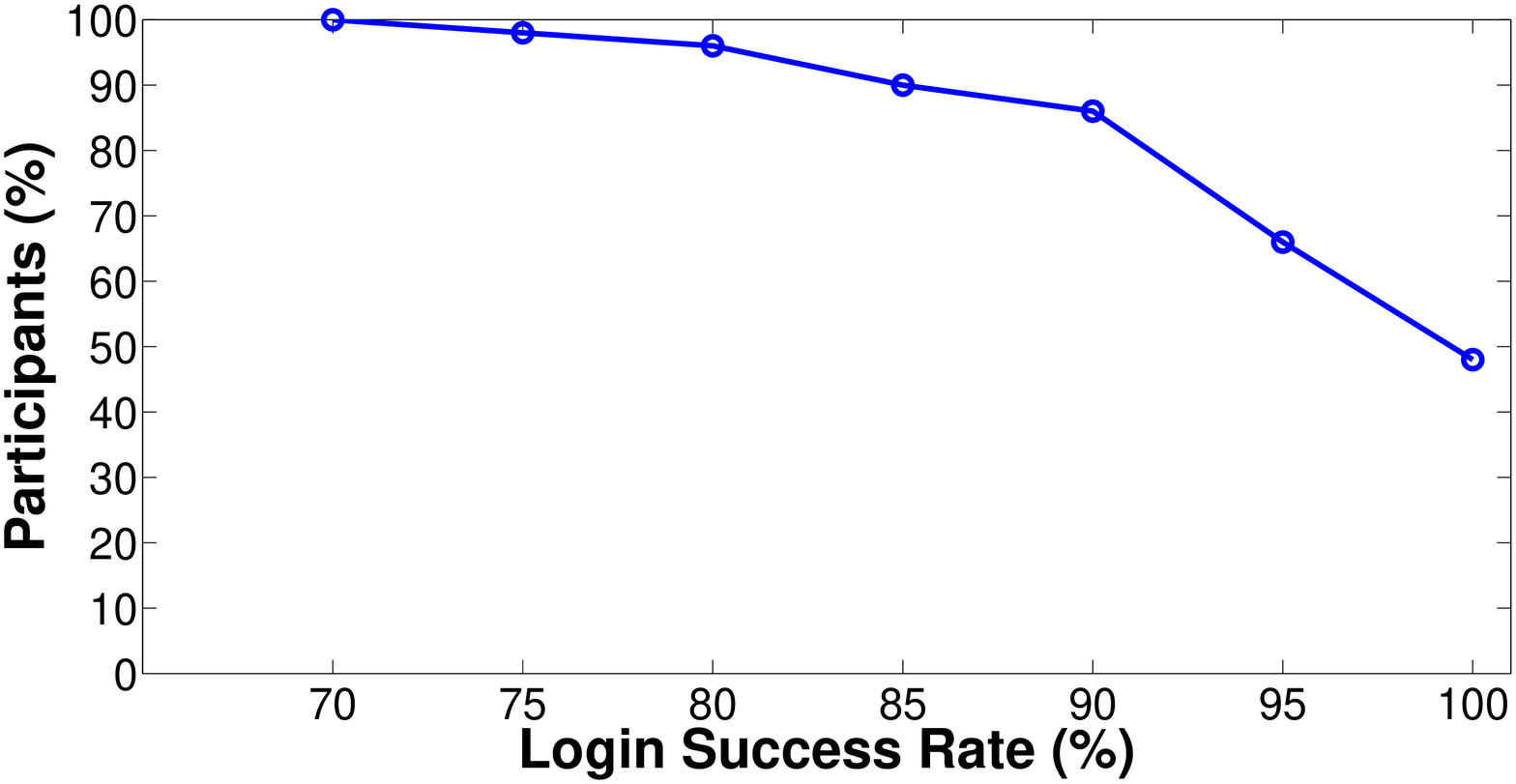}}\hfill
\subfigure[Mean number of attempts]{
\label{fig:attempts}
\includegraphics[width=0.32\linewidth,height=35mm,trim=1.5cm 0 4cm 0]{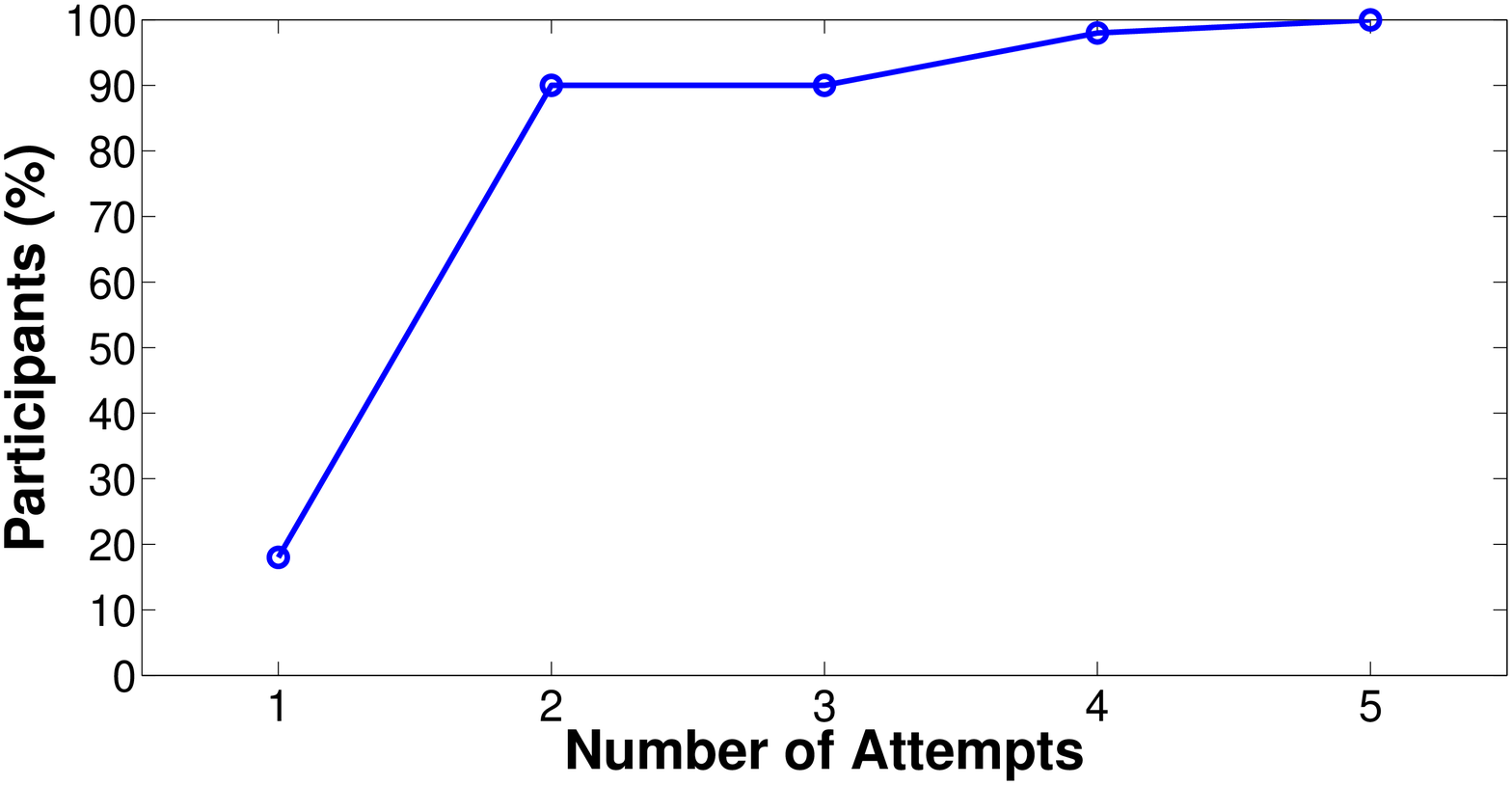}}\hfill
\subfigure[Login time]{
\label{fig:time}
\includegraphics[width=0.32\linewidth,height=35mm,trim=1.5cm 0 4cm 0]{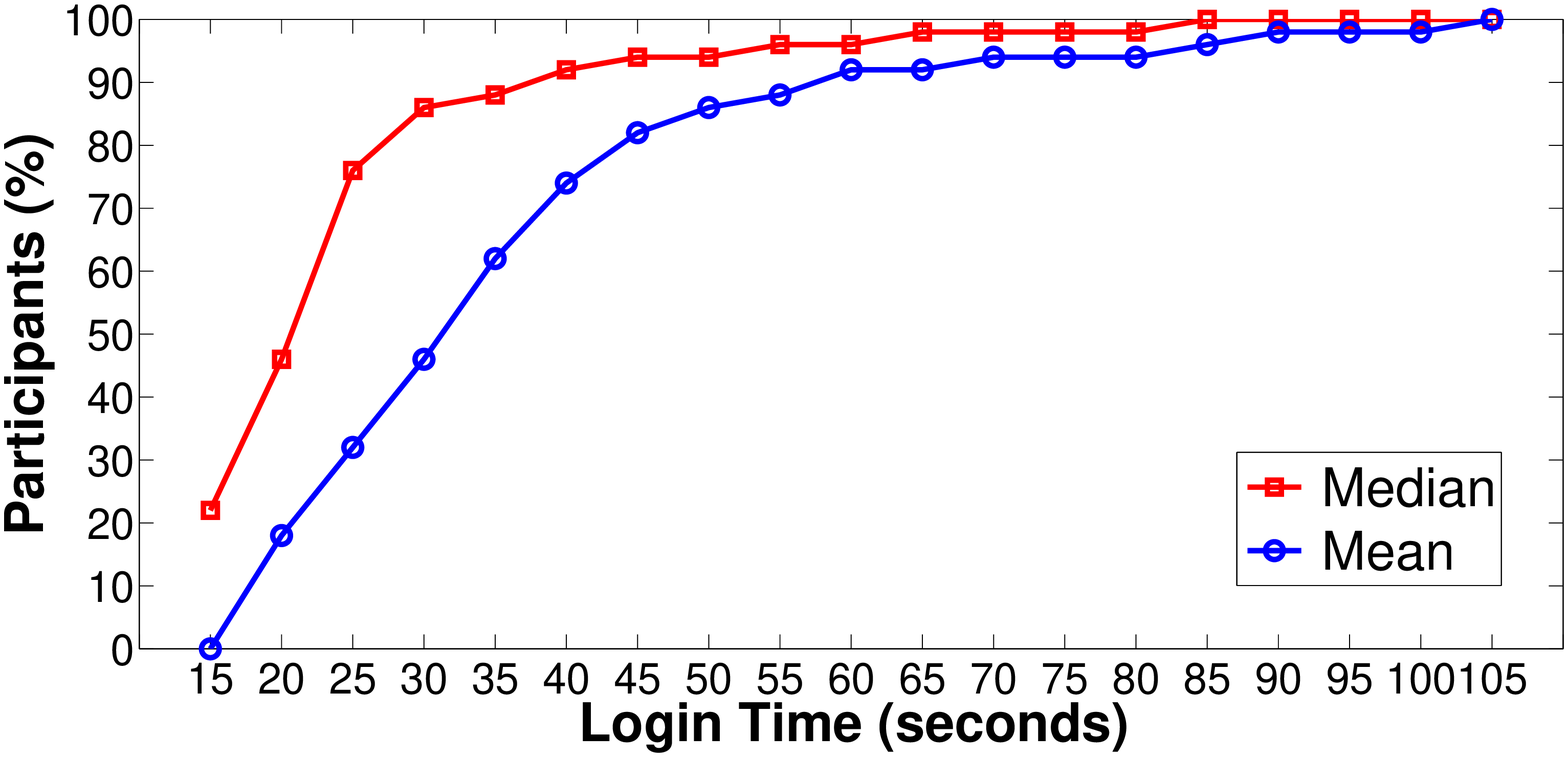}}
\vskip -5 pt
\caption{{\em Field Study:} Login performance (50 participants)}
\label{fig:login}
\vskip -0.3cm
\end{figure*}

\begin{table}[b]
\vskip -5 pt
\renewcommand{\arraystretch}{1.3}
\caption{{\em Field Study:} Registration time (seconds)}  \centering
\vspace{0.1cm}
\begin{tabular}{clccc}
Mean&Med&SD&Max&Min\\ 
\hline
$203$&$154$&$162$&$593$&$25$\\
\end{tabular}
\label{tab:fs_reg}
\vskip -10 pt
\end{table}

\begin{table}[b]
\renewcommand{\arraystretch}{1.3}
\caption{{\em Field Study:} Overall login performance [Login Sessions: \textbf{$1781$}, Login Success Rate: \textbf{$96.1$}\%]}  \centering
\begin{tabular}{c@{\hskip 0.5cm}rrrrr}
&Mean&Med&SD&Max&Min\\ 
\hline
No. of Attempts&$1.3$\phantom{a}&$1$\phantom{a}&$1.3$\phantom{a}&$22$\phantom{a}&$1$\phantom{a}\\
Login Time (sec.)&$32$\phantom{a}&$19$\phantom{a}&$59$\phantom{a}&$997$\phantom{a}&$5$\phantom{a}\\
\end{tabular}
\label{tab:fs_sum}
\end{table}

\subsection{Study Design} 

We conducted the field study on a computer science class with both undergraduate and graduate students. 
Students were informed verbally in lecture that the scheme was not developed by the instructor. We developed a website to let the students access course study materials and their grades on exams and assignments. Upon collecting the students' names from the instructor, a username ({\tt firstname.lastname}) was assigned to each user, and we asked them to create the geopass for their accounts. To protect against unauthorized access, the usernames of the students were pre-stored in the system so that only students in this class could create accounts, one per username. Table~\ref{tab:fs_reg} shows the results for registration, where maximum, minimum, median and standard deviation are cited as \textit{Max}, \textit{Min}, \textit{Med}, and \textit{SD}, respectively.

The GeoPass system was active for $66$ days (Oct. 15 to Dec. 20) during the Fall 2013 semester. Out of $57$ students in this class, $50$ students ($10$ women and $40$ men with a mean age of $24$) gave positive consent to use their login information for the study and signed consent forms before participating in an anonymous paper-based survey at the end of semester. They were compensated with extra credit in a class assignment for participating in this survey, and an alternative assignment was offered for those who did not want to participate.

Our system recorded $1781$ login sessions performed by these $50$ users during the study. The users could log in at anytime from anywhere using their desktop or laptop computers. $56$\% of users reported that in most cases, they logged in from a computer with a $15$-inch screen, while the screen sizes of computers for other users varied from $12$ to $27$ inches. 

During authentication, we started counting login time when the Google map interface was shown to the user after entering her username. A successful attempt required to the user to enter both her username and geopass correctly. An unsuccessful attempt refers only to sessions where the username was correct but the geopass was selected incorrectly.

To reset a geopass, the participant had to send an email to the experimenter from her \textit{.edu} email account, and in response she would receive a link through email to create a new geopass. A few participants had to reset their geopasses within the first few days of the study because of a technical problem in our system. The results reported in this paper do not include any login sessions for these users performed prior to these resets, since the corrupted data had to be deleted while fixing that technical issue. Thereafter, no participant reset her geopass during the study.

\textbf{Limitations.} Our participants were generally young and university educated, so our findings may not generalize to the entire population of Web users.

\subsection{Overall Login Performance}\label{login_field}

In our field study, we recorded $1781$ login sessions, where a single
login session (or {\em login}) by a participant may include multiple
attempts to authenticate successfully. A login is marked as unsuccessful
when the participant leaves the authentication webpage after failing to
click on the correct location. 
To find the full distribution of the number of attempts needed for a successful
login, we did not limit the number of attempts a participant can make
during a login session. One attempt refers to right-clicking at a
location on the Google map.

Participants performed $35.6$ logins on average (minimum: $8$, maximum: $130$). Figure~\ref{fig:num_login} shows the number of logins by the participants, where a $(x,y)$ point represents the percentage of participants ($y$\%) who conducted at least $x$ logins, either successful or unsuccessful.

We measured the average login performance of each participant in her login sessions (see Figure~\ref{fig:login}) and calculated the overall login performances for all of the participants over $1781$ login sessions (see Table~\ref{tab:fs_sum}). The overall login success rate was $96.1$\%. Users required $1.3$ attempts (on average) per successful login, while the average login time was $32$ seconds, with a median of $19$ seconds.

\begin{figure*}[!t]
\centering
\subfigure[Success Rate]{
\label{fig:ch_success}
\includegraphics[width=0.32\linewidth,height=35mm,trim=1.5cm 0 5cm 0]{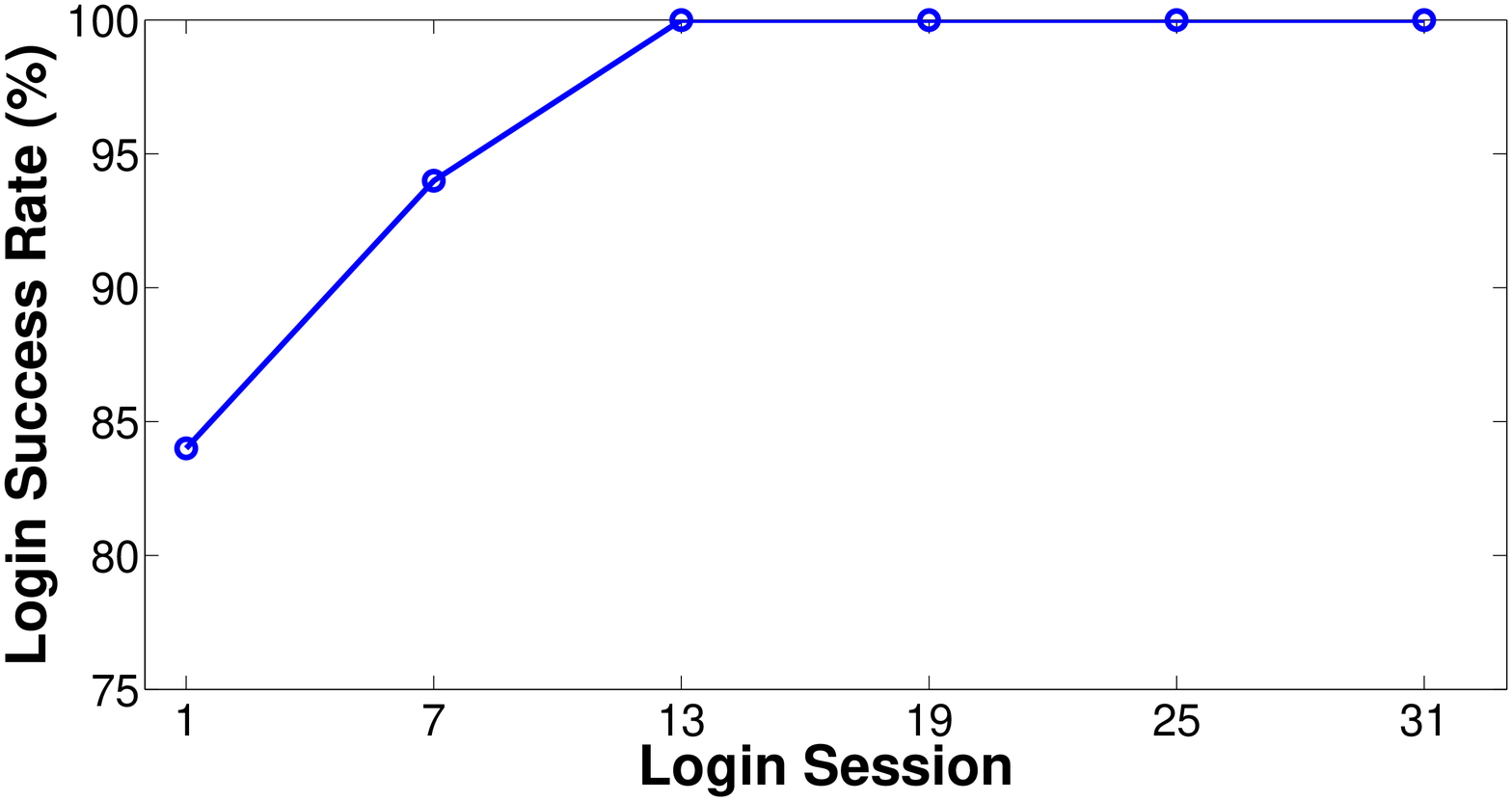}}\hfill
\subfigure[Number of Attempts]{
\label{fig:ch_attempts}
\includegraphics[width=0.32\linewidth,height=35mm,trim=1.5cm 0 5cm 0]{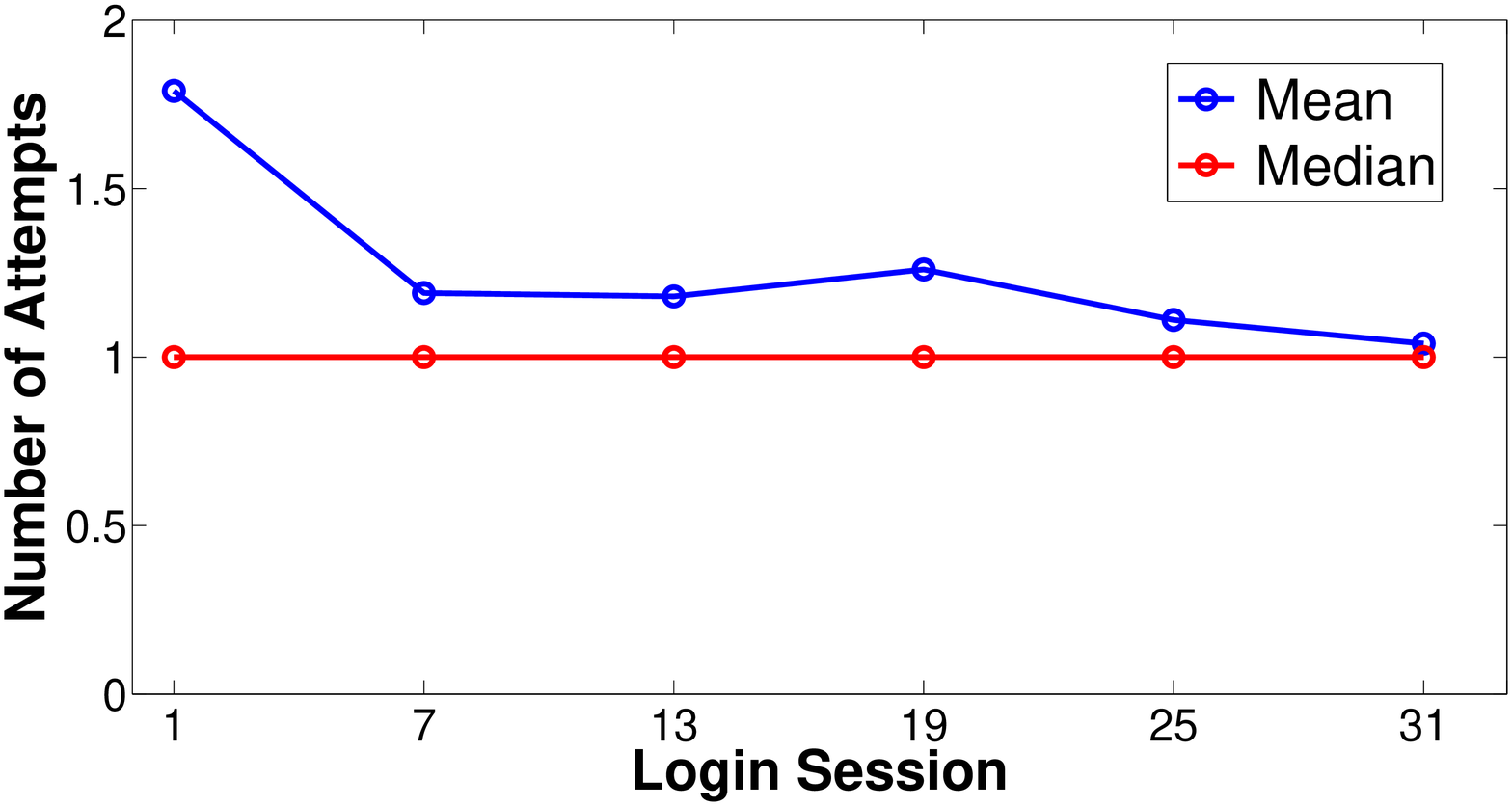}}\hfill
\subfigure[Login Time]{
\label{fig:ch_time}
\includegraphics[width=0.32\linewidth,height=35mm,trim=1.5cm 0 5cm 0]{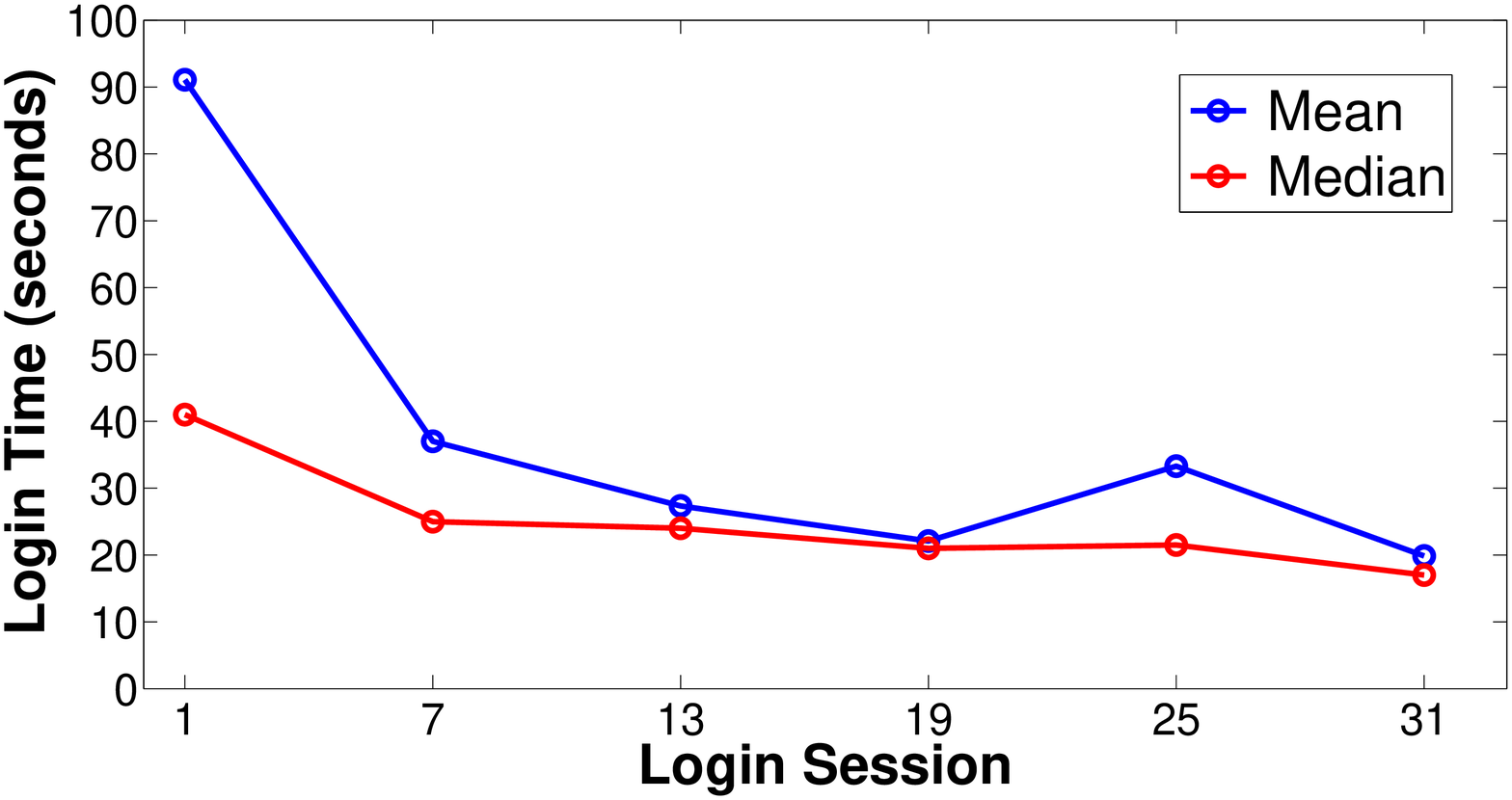}}
\vskip -5 pt
\caption{{\em Field Study:} The change in login performance over the sessions}
\label{fig:ch_login}
\vskip -0.3cm
\end{figure*}

\subsubsection{Login performance distribution among users}

Figures~\ref{fig:success}, \ref{fig:attempts}, and \ref{fig:time} show
empirical CDFs (ECDFs) of login performance statistics taken over the
users in our study (in Figure~\ref{fig:success}, the x-axis is shown
with increasing success rates and thus appears reversed). 


Figure~\ref{fig:success} shows login success rates among
participants. $48$\% of participants had a $100$\% login success rate,
$82$\% had at least a $90$\% success rate, and $96$\% had at least an
$80$\% success rate. The minimum success rate for any participant was
$70$\%.

Figure~\ref{fig:attempts} shows the average number of attempts per successful login among the participants. $90$\% of participants made at most two attempts on average to authenticate successfully, and $100$\% of participants logged in successfully within five attempts on average.

Figure~\ref{fig:time} shows the average login time among
participants. The average login time was $20$ seconds or less for $18$\%
of participants and $30$ seconds or less for $46$\% of participants. The
median login time was $15$ seconds or less for $22$\% of participants
and $30$ seconds or less for $76$\% of participants.

\subsection{Training Effects} \label{training_field}

To determine the extent of any training effects for GeoPass users in a real-world setting, we analyzed the change in login performance over login sessions. The results are shown in Figure~\ref{fig:ch_login}. We illustrate the results at $x^{th}$ login session ($x=1,7,13,19,25,31$), where there are five login sessions between each value of $x$. Here, we consider up to the $31^{st}$ login session, since using the next value of $x$ ($37$) would make for a rather small sample size ($15$ users). For significance tests in this section, we used Wilcoxon signed-rank tests for matched pairs of subjects and Wilcoxon-Mann-Whitney tests for unpaired results.

We note that the sample size changes (shrinks) for each successive value
of $x>7$. As we are looking for a training effect, we may be concerned
about the remaining population of users being more adept at using the
system than those who have stopped logging in. Our results, however,
show that the number of login sessions performed by a participant did
not have a strong correlation with her login success rate ($r=0.18$),
number of attempts for successful login ($r=-0.11$), or login time
($r=-0.21$). As the strongest training effects that we found occurred by
session $x=7$, this is not a surprising result.

A given $(x,y)$ point in a graph in Figure~\ref{fig:ch_login} represents the average login performance ($y$) of the participants calculated over the $x^{th}$ login session of each individual. Note that the $x^{th}$ login session of any given participant likely occurred at a different time than that of other participants. The number of participants varied for different values of $x$ (login session), since the participants performed different numbers of logins (see Figure~\ref{fig:num_login}). Table~\ref{tab:users} represents the number of participants in each of $x^{th}$ login sessions.

\textbf{Login Success rate (Figure~\ref{fig:ch_success}).}  The login success rate was $84$\% in the first login session and $94$\% in the $7^{th}$ login session, which was increased to $100$\% by the $13^{th}$ login session. The login success rate remained at $100$\% in the $19^{th}$, $25^{th}$, and $31^{st}$ login sessions.

\textbf{Number of attempts (Figure~\ref{fig:ch_attempts}).} The mean number of attempts for a successful login was $1.79$ in the first login session, which decreased to $1.19$ in the $7^{th}$ login session. Except for the $19^{th}$ login session, the mean number of attempts for successful logins decreased over the login sessions as shown in Figure~\ref{fig:ch_attempts}. The median number of attempts for successful logins remained constant over the login sessions.

The results for significance tests show that the number of attempts for successful login in the $7^{th}$ ($V=68.5$, $p<0.05$), $13^{th}$ ($W=795.5, p<0.01$), $19^{th}$ ($W=628, p<0.01$), $25^{th}$ ($W=475.5, p<0.01$), and $31^{st}$ ($W=333.5, p<0.01$) login sessions were significantly less than that in the first login session. We did not find significant difference in number of attempts between any other pair ($x,y$) of login sessions ($x,y=1,7,13,19,25,31$). 

\textbf{Login time (Figure~\ref{fig:ch_time}).}  The mean login time was $91$ seconds in the first login session, which decreased to $37$ seconds in the $7^{th}$ login session. As shown in Figure~\ref{fig:ch_time}, mean and median login times decreased over the login sessions, where we find an exception at the $25^{th}$ login session.

The results for significance tests show that login time in the $7^{th}$ ($V=266$, $p<0.001$), $13^{th}$ ($W=430, p<0.001$), $19^{th}$ ($W=245, p<0.001$), $25^{th}$ ($W=228, p<0.001$), and $31^{st}$ ($W=114, p<0.001$) login sessions were significantly less than that in the first login session. We also found that login time in the $13^{th}$ ($W=867.5, p<0.05$), $19^{th}$ ($W=592, p<0.01$), $25^{th}$ ($W=473.5, p<0.01$), and $31^{st}$ ($W=267.5, p<0.001$) login sessions were significantly less than that in the $7^{th}$ login session, and login time in the $31^{st}$ login session was significantly less than that in the $13^{th}$ login session ($W=319.5$, $p<0.05$). No significant difference in login time was found between any other pair $x,y$ of login sessions ($x,y=1,7,13,19,25,31$). 

\begin{table}[t]
\renewcommand{\arraystretch}{1.3}
\caption{{\em Field Study:} Number of participants in the $x^{th}$ login session}  \centering
\begin{tabular}{c@{\hspace{0.7cm}}rrrrrr}
$x$ &$1$&$7$&$13$&$19$&$25$&$31$\\ 
\hline
Participants&$50$&$50$&$44$&$35$&$28$&$21$\\
\end{tabular}
\label{tab:users}
\vskip -0.3cm
\end{table}

\section{Interference Study}\label{interference}
In this section, we discuss the procedures used in our interference study and then provide the study results, which show login performance when using multiple geopasses and the corresponding interference effects.

\subsection{Study Design} 
We conducted the interference study in a course on computing basics. 
Out of $60$ students in this class, $18$ students ($11$ women and $7$ men) participated in our study. Their mean age was $23$. The subjects were compensated with extra credit in a class assignment for participating in this study, and an alternative assignment was offered for those who did not want to participate. No student participated in both field and interference studies.

\begin{table*}[t]
\renewcommand{\arraystretch}{1.3}
\caption{{\em Interference Study:} Login performance of the participants}  \centering
\begin{tabular}{cccc@{\hskip 0.6cm}cccccc@{\hskip 0.4cm}ccccc}
\toprule

&\multirow{2}{*}[-0.2cm]{Accounts} & \multirow{2}{*}[-0.2cm]{Sitting} & Success & \multicolumn{5}{c}{Number of Attempts}  & \hspace{0.1cm}  & \multicolumn{5}{c}{Login Time}             \\ 

\cline{5-9}\cline{11-15}
& &  & Rate (\%) & Mean & Med & SD   & Max & Min & & Mean  & Med & SD    & Max & Min \\
\cline{2-15}
& \multirow{2}{*}{Bank}     & login 1 & $56$  & $5.2$  & $3$      & $7$ & $26$      & $1$       && $99$  & $91$   & $49$ & $188$     & $41$      \\
\cline{3-15}
&      & login 2 & $61$    & $3$    & $2$      & $2.2$ & $7$   & $1$       && $79$ & $60$     & $60$ & $203$     & $12$      \\
\cline{2-15}
Separate&\multirow{2}{*}{Email}    & login 1  & $78$    & $2.1$ & $1$      & $1.9$ & $6$    & $1$  && $51$ & $40$   & $43$ & $146$   & $11$      \\
\cline{3-15}
View&     & login 2  & $78$  & $3$    & $1$   & $4.7$ & $18$  & $1$    && $35$ & $37$   & $21$  & $72$      & $8$  \\
\cline{2-15}
&\multirow{2}{*}{Movie} & login 1  & $56$ & $1.6$  & $1$  & $1.6$ & $6$  & $1$   && $33$  & $28$   & $22$ & $76$  & $9$       \\
\cline{3-15}
&     & login 2    & $72$   & $2.7$ & $1$  & $2.6$ & $8$    & $1$   && $25$ & $24$  & $11$ & $44$      & $10$      \\
\cline{2-15}
&\multirow{2}{*}{Deals}     & login 1    & $44$ & $2.5$  & $1$   & $3.2$ & $10$    & $1$    && $57$ & $36$   & $49$ & $167$  & $10$      \\
\cline{3-15}
&     & login 2   & $56$   & $4$    & $1.5$    & $4.8$ & $16$   & $1$   && $47$  & $35$  & $30$ & $99$      & $14$     \\
\hline
\multicolumn{2}{c}{Combined}     & login 1 & $58$ & $2.8$ & $1$  & $4.2$ & $26$   & $1$   && $59$ & $43$   & $47$ & $188$     & $9$          \\
\cline{3-15}
\multicolumn{2}{c}{View}& login 2 & $67$   & $3.1$ & $1$   & $3.7$ & $18$  & $1$    && $45$ & $35$   & $39$    & $203$     & $8$      \\
\bottomrule
\end{tabular}
\label{tab:is_sum}
\vskip -0.3cm
\end{table*}

\subsubsection{Procedure} The three-week-long interference study included three sessions, which we will call {\em sittings}, each held one week apart. The class included a two-hour lab session each week, where each student used a desktop computer meant for exercises with common productivity software packages. The lab exercises could be done relatively easily for most students, leaving time for us to conduct sittings during the lab sessions. Each sitting took no more than $30$ minutes.

Haque et al.~\cite{hierarchy} classify websites into four categories: i) financial (e.g. WellsFargo.com), ii) identity (e.g. Gmail), iii) content (e.g., Netflix, Weather.com), and iv) sketchy (unfamiliar sites offering coupons and often attracting transient user relationships). We built one website from each of the above categories and refer to them in this paper as \textit{bank}, \textit{email}, \textit{movie}, and \textit{deals}, respectively. Each site was equipped with GeoPass for user authentication and could be accessed from any computer through the Internet. The sites were designed to have the images and layouts from familiar commercial sites, with the exception of the deals site, which was designed to look more questionable.

Before the first sitting, we contacted students in the class through email, giving them an overview of our study and asking them about their interest to take part in the study. $18$ students responded positively. We then collected their names and established a username ({\tt firstname.lastname}) for each participant to be used for their four accounts. 

In the first sitting, we gave the participants an overview of GeoPass and asked them to create a location-password for each account. To best study interference effects, the participants were asked to create a distinct geopass for each account. In the second and third sittings, users were asked to log into their four accounts from the lab computers. We refer to these sittings as \textit{login 1} and \textit{login 2}, respectively. The participants could log into the sites in any order. We conducted an anonymous paper-based survey at the end of third sitting.

If a participant failed to log into an account after five attempts, she was shown a button that she could use to view her geopass. She was also allowed to make more attempts without viewing her location-password. Once the button was clicked to view the geopass, however, the participant was no longer able to attempt to log into that account for that sitting.

\subsubsection{Limitations} As the study was performed in a lab setting, we were only able to gather data from $18$ participants. The participants were young and university educated, which may not generalize to the entire population. 

We followed the method of Chiasson et al.'s extensive study on interference~\cite{interference4}, in which all the passwords were created in the same sitting. This means, however, that registration is not similar to real life, in which the geopasses would likely be created over time and possibly in different contexts, e.g. in different rooms or with different computers. Addressing all of these issues would be challenging for such a study.

\subsection{Login Performance}\label{login_interference} 
Each of the 18 participants logged into four accounts in both \textit{login 1} and \textit{login 2}, making a total of $72$ login sessions in each sitting. We examine the login performance both for each of the four account types separately (the {\em separate view}) and for all four account types combined (the {\em combined view}). The results for login performances are shown in Table~\ref{tab:is_sum}.  

\textbf{Login Success rate.} In the combined view, the overall login success rates were $58$\% in \textit{login 1} and $67$\% in \textit{login 2} (see Table~\ref{tab:is_sum}). We use a McNemar's test to compare login success rate between \textit{login 1} and \textit{login 2}, since we get matched pairs of subjects in this case. Our analysis shows that login success rate did not differ significantly between \textit{login 1} and \textit{login 2} in either the combined view or the separate view. 


\textbf{Number of attempts.} The mean number of attempts for successful logins were $2.8$ in \textit{login 1} and $3.1$ in \textit{login 2} for combined view, while the median was $1$ in both sittings (see Table~\ref{tab:is_sum}). We do not get matched pairs of subjects while comparing two sittings in terms of login time or number of attempts for successful logins, since some participants who logged in successfully in one sitting failed in the other sitting. So we use a Wilcoxon-Mann-Whitney test to evaluate the differences in number of attempts for successful logins between \textit{login 1} and \textit{login 2}. 

In the combined view, we did not find a significant difference in the number of attempts for successful logins between \textit{login 1} and \textit{login 2}. However, in the separate view for bank accounts, the number of attempts for successful logins in \textit{login 1} was significantly higher than that in \textit{login 2} ($W = 110$, $p < 0.01$). For the other account types, the differences between \textit{login 1} and \textit{login 2} were not significant.


\textbf{Login time.} In the combined view, the mean times for successful logins were found to be $59$ seconds in \textit{login 1} and $45$ seconds in \textit{login 2}, while the medians were $43$ seconds in \textit{login 1} and $35$ seconds in \textit{login 2} (see Table~\ref{tab:is_sum}). We use a Wilcoxon-Mann-Whitney test to evaluate the differences in time for successful logins between \textit{login 1} and \textit{login 2}. We did not find any significant difference in the time for successful logins between \textit{login 1} and \textit{login 2} in either the combined view or the separate view.


\begin{table*}[t]
\renewcommand{\arraystretch}{1.3}
\caption{Summary of the interference effect}  
\centering
\begin{tabular}{clcccccc}
\toprule
\multicolumn{2}{c}{\multirow{3}{*}[-0.3cm]{Accounts}} & \multirow{3}{*}[-0.3cm]{Sitting} &  &  & \multicolumn{3}{c}{Failed Attempts (\%)} \\             
\cline{6-8}
 & & &  Total &  Successful & \multicolumn{2}{c}{Interference} & \multirow{2}{*}{\hskip 0.3cm Non-Interference} \\   
\cline{6-7}                                                                              
 & & & Attempts & Attempts (\%) & Accurate      & Non-Accurate     &           \\ 
\hline
\hspace{0.2cm} & \multirow{2}{*}{Bank}   & login 1    & $101$ & $9.9$   & $7.9$ & $39.6$ & $42.6$ \\   
\cline{3-8}                                                                                               
& & login 2 & $85$  & $12.9$   & $0$  & $48.2$  & $38.8$  \\                                              
\cline{2-8} 
& \multirow{2}{*}{Email} & login 1 & $50$ & $28$    & $2$  & $32$     & $38$     \\
\cline{3-8}
& \multicolumn{1}{c}{} & login 2  & $67$   & $20.9$  & $0$ & $29.9$ & $49.3$   \\
\cline{2-8}
& \multirow{2}{*}{Movie}  & login 1  & $55$ & $18.2$   & $1.8$  & $49.1$    & $30.9$   \\   
\cline{3-8}
& \multicolumn{1}{c}{} & login 2  & $61$ & $21.3$    & $1.6$    & $36.1$    & $41$ \\   
\cline{2-8}
& \multirow{2}{*}{Deals} & login 1  & $76$ & $10.5$  & $5.3$     & $38.2$   & $46.1$   \\ 
\cline{3-8}
& \multicolumn{1}{c}{} & login 2  & $96$  & $10.4$     & $5.2$      & $30.2$    & $54.2$ \\ 
\cline{2-8}
& \multirow{2}{*}{Overall} & login 1  & $282$  & $14.9$ & $5$   & $39.8$ & $40.4$ \\ 
\cline{3-8}
& \multicolumn{1}{c}{} & login 2 & $309$ & $15.5$  & $1.9$  & $36.3$  & $46.3$\\
\bottomrule
\end{tabular}
\label{tab:interference}
\vskip -0.3cm
\end{table*}


\subsection{Interference Effect}\label{interference_effect} 
In each sitting, every participant was asked to complete four login sessions, each for one account. We refer to the account corresponding to current login session as the \textit{visible account} and refer to the other three accounts as \textit{invisible accounts}. For example, when a participant attempts to log into the bank account, the bank account is visible, while email, movie, and deals accounts are considered invisible. Thus, a successful login requires the user to select the geopass of the visible account. Because of interference effect, a user may make mistakes and click on the geopass of an invisible account. Table~\ref{tab:interference} shows the summarized results for interference effects in our study. 

We did not restrict the number of attempts a participant can make for a successful login, and clicking at a location other than the geopass of the visible account results in an unsuccessful attempt. We figure out the impact of interference on the failure of an attempt in the following way: For each unsuccessful attempt, we measure the distances (in kilometers) between the clicked location and her geopasses for each of the four accounts. In this way, we find the account whose geopass is closest to the clicked location. If the closest account is the visible account, we assume that interference did not impact the failed attempt, and we show this as \textit{non-interference} in Table~\ref{tab:interference}. If the closest account is an invisible account, we say that the attempt fails because of the interference effect. In this case, if the clicked location is a correct geopass for the invisible account, we classify it as \textit{accurate interference}, and otherwise we call it \textit{non-accurate interference}. 

Our results (see Table~\ref{tab:interference}) show that $14.9$\% of $282$ attempts succeeded in \textit{login 1}, while $44.8$\% attempts failed because of interference effects (considering both accurate and non-accurate interferences). In \textit{login 2}, out of $309$ attempts $15.5$\% were successful, and $38.2$\% attempts failed because of interference effects.

We used a Wilcoxon signed-rank test to evaluate the difference between \textit{login 1} and \textit{login 2} in terms of the number of failed attempts because of interference effects, as we have matched pairs of subjects in our dataset. Considering both accurate and non-accurate interferences, we didn't find any significant difference between \textit{login 1} and \textit{login 2} in the number of failed attempts due to interference effects ($V = 422.5$, $p = 0.72$). No significant difference was found either between \textit{login 1} and \textit{login 2} if we consider only accurate interference ($V = 14.5$, $p = 0.36$) or only non-accurate interference ($V=369$, $p=0.98$). 

\begin{table}[b]
\renewcommand{\arraystretch}{1.3}
\caption{{\em Interference Study:} User Behavior in creating multiple location-passwords}
\centering
\begin{tabular}{@{}c@{\hskip 0.25cm}cc@{}c@{}cccc@{}}
& \multicolumn{2}{c}{Same Country} & \hspace{0.3cm} & \multicolumn{4}{c}{Same City} \\
\cline{2-3} \cline{5-8} 
\# geopasses &\hskip 0.2cm $4$& $3$ & & $4$     & $3$     & $2$     & None  \\ 
\hline
Users (\%) &\hskip 0.2cm $89.9$ & $11.1$  & & $27.8$ & $16.7$ & $11.1$ & $44.4$ \\
\end{tabular}
\label{tab:is_city}
\end{table}

\subsection{Correlation between Distance and Interference Effects}\label{distance_interference} 

\begin{figure}[t]
\centering
\includegraphics[width=91mm]{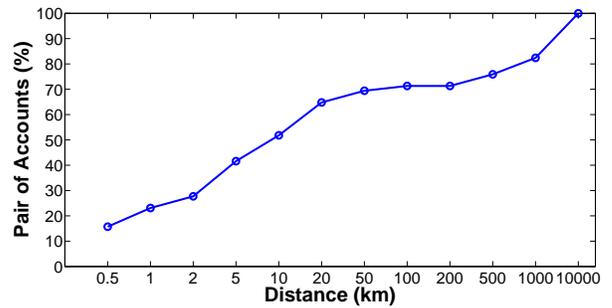}
\caption{{\em Interference Study:} Distances between geopasses for each pair of accounts}
\label{fig:is_dist}
\end{figure}

We examined the city~\footnote{The {\em city} can mean a city, town, or smaller jurisdiction, and different towns in the same metropolitan area will be considered distinct.} of each geopass as indicated by the Google Maps API and found that $27.8$\% of participants chose all four geopasses in the same city, while $16.7$\% of participants created three location-passwords in the same city and one in a different city (see Table~\ref{tab:is_city}). We also measured the distance (in kilometers) between each pair of geopasses of a participant. Since each participant created four location-passwords, we had six pairs of accounts for each participant and $108$ pairs in total for $18$ participants. Figure~\ref{fig:is_dist} shows an empirical CDF (ECDF) for the distances between geopasses of each pair of accounts, where over one-quarter ($27.8$\%) of geopass pairs were within $2.0$ km.

It is possible that a user who selects two geopasses near each other may confuse them, leading to interference effects. We thus seek to determine whether the interference effect between a pair of accounts had any correlation with the distance between corresponding pairs of geopasses.

To measure the interference effect between two accounts for a given participant, such as bank and email, we counted the number of failed attempts $T_1$ for which bank was the visible account and email was the invisible account with the closest geopass (compared to the geopasses of the movie and deals accounts) to the clicked location. We then calculated $T_2$ representing the number of unsuccessful attempts for which email was the visible account and bank was the invisible account with the closest geopass to the clicked location. Here, $T = T_1+T_2$ refers to the interference effect between bank and email accounts for this participant. We also measured the distance $D$ (in kilometers) between the location-passwords for the bank and email accounts of the given participant. In this way, we measured $D$ and $T$ for each pair of accounts of every participant and then calculated the correlations between $D$ and $T$. However, we didn't find any strong correlations in this respect either in \textit{login 1} ($r=0.12$) or in \textit{login 2} ($r=0.02$). 

We also considered accurate interference to find correlations with the distances between pairs of geopasses. The process of calculation was same as described above except for $T$, which now represents the number of attempts where the clicked location accurately represents the location-password of the corresponding invisible account (email or bank in the above example). In this respect, no strong correlation was found either in \textit{login 1} ($r=0.02$) or in \textit{login 2} ($r=0.0002$). Also, for non-accurate interference, we did not find any strong correlation: \textit{login 1} ($r=0.12$), \textit{login 2} ($r=0.02$). 

Thus, the participants did not appear to be confused by geopasses that were close together on the map.

\begin{table}[b]
\renewcommand{\arraystretch}{1.3}
\caption{{\em Field and Interference Studies:} Responses to the Question: ``Which of the following locations corresponds to your geopass?" [Cur: Current, Prev: Previous, Bk: Bank, Em: Email, Mv: Movie, Dl: Deal, wp: workplace]}
\centering
\begin{tabular}{l@{}r@{\hskip 0.5cm}c@{\hskip 0.6cm}cccc}\toprule
\multicolumn{2}{c}{\multirow{3}{*}[-0.5cm]{Locations}}& \multicolumn{5}{c}{Participants (\%)}\\ 
\cline{3-7}
\hspace{0.2cm} & & Field& \multicolumn{4}{c}{Interference Study}\\ 
\cline{4-7}
&  & Study & Bk & Em & Mv & Dl \\
\hline
& Cur/Prev home & $34.7$ & $22.2$& $22.2$ & $27.8$ & $16.7$\\ 
\cmidrule{2-7}
& Cur/Prev school/wp & $10.2$ & $33.4$& $27.8$ & $11.1$ & $16.7$\\ 
\cmidrule{2-7}
& Place of birth  & $0$ & $0$& $0$ & $11.1$ & $0$\\
\cmidrule{2-7}
& A visited place & $6.1$ & $5.6$& $22.2$ & $16.7$ & $16.7$\\ 
\cmidrule{2-7}
& My favorite place  & $22.5$ & $11.1$& $5.6$ & $5.6$ & $27.8$\\ 
\cmidrule{2-7}
& A historical place  & $4.1$ & $0$& $0$ & $0$ & $0$\\ 
\cmidrule{2-7}
& An unusual place & $10.2$ & $5.6$ & $0$ &$5.6$ &$0$\\ 
\cmidrule{2-7}
& None of the above & $12.2$ & $22.2$& $22.2$ & $22.2$ & $22.2$\\ 
\bottomrule
\end{tabular}
\label{tab:place}
\vskip -0.3cm
\end{table}

\begin{table*}[t]
\renewcommand{\arraystretch}{1.3}
\caption{User Feedback on GeoPass. Scores are out of 10. * indicates that the scale was reversed}  \centering
\begin{tabular}{@{}lcccccccccc@{}}
\toprule
\multirow{2}{*}[-0.15cm]{\hspace{2cm}Questions}& \multicolumn{4}{c}{Field Study} &\hspace{0.1cm}& \multicolumn{4}{c}{Interference Study}&\hspace{0.1cm}\\
\cline{2-5}\cline{7-11}
 & Mode&Med&Mean&SD&&Mode&Med&Mean&SD\\ 
\hline
 I could easily sign up with GeoPass&$9$&$9$&$8.6$&$1.2$&&$10$&$8.5$&$7.7$&$2.9$\\
\cline{1-11}
 The login using GeoPass was easy&$10$&$9$&$8.3$&$1.6$&&$10$&$8$&$7.7$&$2.6$\\ 
\cline{1-11}
 GeoPass is easy to remember &$8$&$8$&$8.3$&$1.3$&&$1$&$6.5$&$5.6$&$3.1$\\
\cline{1-11}
*I found GeoPass too time-consuming &$8$&$7.5$&$5.9$&$2.4$&&$7$&$7$&$5.7$&$2.4$\\
 (i.e., I found GeoPass less time consuming) & & & & & & & &\\ 
\cline{1-11}
With practice I could quickly enter geopass&$10$&$9$&$8.6$&$1.5$&&$10$&$8$&$7.6$&$2.5$\\ 
\cline{1-11}
*I prefer user-chosen text passwords to GeoPass &  \multirow{2}{*}{$6$} & \multirow{2}{*}{$5$} & \multirow{2}{*}{$3.9$} & \multirow{2}{*}{$2.4$}&& \multirow{2}{*}{$1$} & \multirow{2}{*}{$4.5$} & \multirow{2}{*}{$3.3$} &\multirow{2}{*}{$2.8$}\\
 (i.e. I prefer GeoPass to user-chosen text passwords) &&&&&& \\
\bottomrule
\end{tabular}
\label{tab:feedback}
\vskip -0.3cm
\end{table*}

\subsection{Banking Geopass Selection}\label{spec} 

In real life, it would be difficult to restrict users from choosing their current or previous home, school, or work locations as geopasses. Instead, it would be realistic to suggest for security reasons that choosing such places may make their geopasses vulnerable to targeted guessing attacks. So we verbally gave this simple security suggestion to the participants at the beginning of the study but did not restrict them from choosing their home, school, or workplace as location-passwords. 

In the post-experiment questionnaire at interference study, we asked participants, ``Which of your four geopasses is the strongest one against guessing attack?". In response, most of the participants ($61$\%) picked their geopass for the bank account. We also found that for the majority of participants ($55.6$\%), geopasses for bank accounts refer to their \textit{current/previous home}, \textit{schools}, or \textit{workplaces} (see Table~\ref{tab:place}), which are generally more vulnerable to guessing attacks than geopasses that do not represent a publicly-known location (e.g., \textit{My favorite place}). 

To explain these apparently conflicting results, we could say that the participants did not carefully judge the strength of their geopasses. However, we also offer an alternative explanation.

First, we note some additional results from the interference study. For each sitting (e.g., \textit{login 1}, \textit{login 2}) in the study, we compared each pairs of accounts in terms of login success rate (McNemar's test), number of attempts for successful login (Wilcoxon-Mann-Whitney test), and login time (Wilcoxon-Mann-Whitney test). The appropriateness of using McNemar's test and Wilcoxon-Mann-Whitney test are explained in~\S\ref{test}. For login success rate, we did not find any significant difference between any pair of accounts. However, participants required significantly higher login times for bank accounts in comparison to other accounts in \textit{login 1}: email ($W = 110$, $p < 0.05$), movie ($W = 92.5$, $p < 0.01$), deal ($W = 66$, $p < 0.05$), and in \textit{login 2}: email ($W = 114$, $p < 0.05$), movie ($W = 118.5$, $p < 0.01$). The number of attempts was also significantly higher for bank accounts than email ($W = 105.5$, $p < 0.05$) and movie ($W = 84.5$, $p < 0.01$) accounts in \textit{login 1}. 

Since bank passwords are required to be both memorable and secure, participants might have chosen geopasses that refer to highly memorable locations and then picked less obvious points around that location to increase the security of their authentication secrets. We speculate that such a strategy might have led to the higher login times for bank accounts, as it is generally easier and less time-consuming to find a clear landmark on the digital map. 

We note that we did not have a way to validate this explanation through our anonymous survey.

\section{User Feedback} \label{feedback}

To gain an understanding of users' perceptions of GeoPass, we asked users to answer Likert-scale questions at the end of both the interference and field studies. We used ten-point Likert scales, where anchors were included on the ends of the scale ($1$ indicating \textit{strong disagreement} and $10$ equaling \textit{strong agreement} with the given statement). We reversed some of the questions to avoid bias. The scores marked with (*) were reversed before calculating modes, medians, and means. For all questions, a higher score indicates a more positive result for GeoPass. 

Since Likert-scale data are ordinal, it is most appropriate to calculate mode and median for Likert-scale responses. For interested readers, we also report the mean. We say that the user feedback was \textit{positive} when the mode and median values were higher than neutral. We did not get multiple modes in any case.

\textbf{Field Study.} In the field study, where each participant had to remember a single geopass, participants showed high overall satisfaction on the usability (e.g., memorability, ease of registration and login) of GeoPass (see Table~\ref{tab:feedback}). We also found positive feedback about the login time of GeoPass. However, user feedback did not reflect a strong preference for GeoPass to traditional textual passwords. 

\textbf{Interference Study.} In the interference study, participants had to remember multiple geopasses and the majority of participants strongly disagreed with the notion that GeoPass offered good memorability (see Table~\ref{tab:feedback}). The responses for other usability aspects (e.g., ease of registration and login) were found to be positive overall. The majority of participants in the interference study showed strong preference for textual passwords.

We do not make a direct comparison between these two studies because of the differences in study design. 

\section{Shoulder-Surfing Study}\label{shoulder_surfing}
In this section, we discuss the procedure and results of our study of shoulder surfing in GeoPass. We seek to understand whether shoulder surfing would be possible to steal users' login credentials and what aspects of shoulder surfing against GeoPass are particularly easy or difficult.

\subsection{Study Design} In this study, we recruited $30$ participants ($7$ women, $23$ men) from the \textit{Computer Security Club}, a \textit{Computer Security Class}, and a \textit{Computer Security Research Lab} at our university. Their mean age was $23$. The participants were divided into two groups (described below), with $15$ participants in each group. The $i^{th}$ and $(i+1)^{th}$ participants were assigned to different groups.

\subsubsection{Procedure} In our study, the experimenter had a one-on-one session with each user. Written consent was obtained from each participant before the beginning of study. We showed and explained GeoPass to the participants and gave an overview of the shoulder-surfing study. In this study, a participant played the role of a {\em hacker} and the experimenter played the role of a {\em victim}, where the experimenter logged in as a user using a desktop computer with a $15$-inch monitor.

The participants were divided into two groups according to the method of navigation used by the experimenter to find the location passwords: the {\em Panning Group} and the {\em Typing Group}. In the {\em Panning Group}, the experimenter typed the name of the city in the search bar and then used panning for navigation to the location-password. For the {\em Typing Group}, the experimenter typed the full address of the geopass in the search bar for direct navigation to that location without using panning. 

In both groups ({\em Panning} and {\em Typing}), every participant was asked to perform shoulder surfing twice. For one shoulder-surfing attempt, the participant was given a pen and paper to take notes, and for the other shoulder-surfing attempt, she was provided with a $7$-inch \textit{A13 Google Android tablet} with the Google map interface (at \url{http://www.maps.google.com}) open for them to mimic the actions of the experimenter on Google map to gain his location-password. We attempted to randomize the order of pen and paper versus tablet, but due to experimental artifacts, the majority of participants used the tablet first before using pen and paper.

Thus, we get four conditions: the {\em Panning Group} using pen and paper (\textit{Pan-P\&P}), the {\em Panning Group} using the tablet (\textit{Pan-Tab}), the {\em Typing Group} using pen and paper (\textit{Typ-P\&P}), and the {\em Typing Group} using the tablet (\textit{Typ-Tab}). 

For each participant, we used two different geopasses: one in Sydney, Australia (for pen and paper) and another one in Cape Town, South Africa (for the tablet). We expect that most participants would not be familiar with either location.

The experimenter clicked geopasses at zoom level 16 in all logins and maintained an average login time of $35$ seconds for the first group and $28$ seconds for the second group. Note that the mean login time for GeoPass was $32$ seconds in the field study (see Table~\ref{tab:fs_sum}). We let the participants get comfortable using Google maps on the tablet before the start of the study. We hypothesized that the tablet would be easier to use to get precisely the right location.  

During shoulder surfing, the participants could stand behind the experimenter, move to any side, or sit next to him to gain the geopass. Then they were asked to log in using the information they had gained through shoulder surfing. For each participant, the second shoulder-surfing attempt started after she had completed login attempts with the credentials gained through the first shoulder-surfing attempt. Each participant was allowed to make a maximum $10$ login attempts. In the post-experiment anonymous paper-based survey, they were asked about the ease of shoulder surfing on GeoPass. They were compensated with a five dollar gift card for participating in this study.

\begin{table*}[t]
\renewcommand{\arraystretch}{1.3}
\caption{{\em Shoulder-Surfing Study:} Login Performance}  \centering
\begin{tabular}{ccccccccccccccc}\toprule
\multicolumn{2}{c}{\multirow{2}{*}[-0.2cm]{Condition}} & \multirow{2}{*}[-0.2cm]{Material} & Success & \multicolumn{5}{c}{Number of attempts}   &  & \multicolumn{5}{c}{Login Time}             \\ 
\cline{5-9}\cline{11-15}
\hspace{0.2cm} & & & Rate (\%) & Mean & Med & SD & Max & Min & \hspace{0.2cm} & Mean  & Med & SD    & Max & Min \\
\hline
& \multirow{2}{*}{Pan}     & P\&P & $60$  & $1.4$  & $1$  & $0.7$ & $3$      & $1$ & & $77$  & $73$   & $57$ & $176$     & $25$      \\
\cline{3-15}
&      & Tab & $33$    & $2$    & $1$      & $1.4$ & $4$   & $1$  &  & $72$ & $34$     & $74$ & $197$     & $14$      \\
\cline{2-15}
& \multirow{2}{*}{Typ}    & P\&P  & $67$    & $2.4$ & $1$  & $3$ & $10$    & $1$ & & $124$ & $28$   & $304$ & $987$   & $15$      \\
\cline{3-15}
&  & Tab  & $33$  & $2.4$    & $2$   & $1.5$ & $4$  & $1$ & & $88.8$ & $63$   & $62$  & $196$      & $41$  \\
\cline{2-15}
& \multicolumn{2}{c}{Overall} & $48$   & $2$    & $1$ & $1.9$ & $10$   & $1$ & & $94$  & $40$  & $180$ & $987$      & $14$     \\
\bottomrule
\end{tabular}
\label{tab:ss_sum}
\vskip -0.1cm
\end{table*}

\subsubsection{Limitations} Since shoulder surfing is performed by people who intend to steal authentication secrets, possibly professional hackers, recruiting a participant group representing the true population is very difficult. However, our participants satisfied the requirements of Tari et al.~\cite{shoulder06} who stated that students with reasonable background on computer security and authentication systems can be representative of ``potential shoulder surfers" in a lab study. We note that for comparisons between pen and paper and tablet conditions that the majority of participants used the tablet first.

\subsection{Login Performance}\label{shoulder_login} 
We consider a shoulder-surfing attempt to be successful when the participant is able to log in successfully with the location-password that she gains through shoulder surfing. Table~\ref{tab:ss_sum} illustrates the results of login performance of the participants. 

\subsubsection{Success rate} The participants in the Typ-P\&P condition were the most successful with a $67$\% login success rate, while the participants in the Pan-P\&P condition attained a $60$\% success rate. In both the Pan-Tab and Typ-Tab conditions, the login success rate was $33$\%. The overall success rate (considering all conditions) of the shoulder surfers was $48$\%.

Login success rates are a binary measure. We divide the pairs into a group for which we have paired
results (Pan-P\&P:Pan-Tab and Typ-P\&P:Typ-Tab) and a group for which results are not paired (Pan-P\&P:Typ-P\&P, Pan-P\&P:Typ-Tab, Pan-Tab:Typ-Tab, and Typ-P\&P:Pan-Tab). For the paired results, we
perform McNemar's tests, while for the unpaired results, we use chi-squared tests. We found no significant differences between any pair of conditions. How a user navigated to the location-password, by panning or typing, did not impact the success rate of the attack. 

Considering that the success rate was higher for pen-and-paper conditions, though not statistically significant, we conclude that our hypothesis that the tablet would be easier to use for shoulder surfing is not supported by the participants' performances. We discuss this point further in \S\ref{shoulder_feedback}.

\subsubsection{Number of attempts} In our shoulder-surfing study, the mean number of attempts required for a successful login was no greater than $3$ in any condition, and the median was $1$ in all conditions except for Typ-Tab, in which case it was $2$. In the Typ-P\&P condition, one participant required $10$ attempts to log in successfully. No other participant in any condition required more than $4$ attempts for a successful login.

We do not get matched pairs of subjects while comparing login times or numbers of attempts for successful logins in the two study conditions, as a participant might succeed in one condition and fail in
another. Thus, we used a Wilcoxon-Mann-Whitney test and found no significant differences between any pair of conditions. 

\subsubsection{Login time} The mean and median times for successful logins were no less than $70$ seconds and $28$ seconds, respectively, in any condition. One participant required $987$ seconds, which was the maximum login time in any condition, while the minimum login time for any participant was $14$ seconds.

For each pair of conditions, we used a Wilcoxon-Mann-Whitney test to evaluate the differences in time required for successful logins. We found that login time for the Typ-P\&P condition was significantly less than that for the Typ-Tab condition ($W=6$, $p<0.05$), though again we note the limitation of ordering between the tablet and pen and paper conditions. No significant difference in login time was found between any other pair of conditions. 

\subsection{Min-Distances for Unsuccesful Logins}\label{shoulder_distance}

\begin{figure}[t]
\centering
\hskip -0.6cm
\includegraphics[width=91mm]{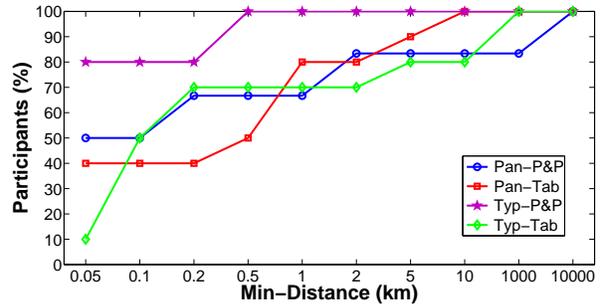}
\vskip -5 pt
\caption{{\em Shoulder-Surfing Study:} Minimum distances between guessed locations and actual geopasses for failed logins}
\label{fig:ss_distance}
\vskip -5 pt
\end{figure}

\begin{table*}[t]
\renewcommand{\arraystretch}{1.3}
\caption{{\em Shoulder-Surfing Study:} Questionnaire responses. Scores are out of 10}  \centering
\begin{tabular}{cl@{\hskip 0.5cm}ccccc}
\toprule
\multicolumn{2}{c}{Question}&Mode&Med&Mean&SD&Significance Tests\\ 
\hline
\hspace{0.2cm}&It was very easy to record the person's geopass with pen and paper &$8$&$8.5$&$8.2$&$1.7$&\multirow{2}{*}{$V=304.5$, $p<0.01$}\\
\cline{2-6}
&It was very easy to record the person's geopass with a tablet &$4$&$5.5$&$5.3$&$2.7$&\\
\cline{2-7}
&It was very easy to select the location from recorded  &\multirow{2}{*}{$10$}&\multirow{2}{*}{$8$}&\multirow{2}{*}{$7.9$}&\multirow{2}{*}{$2.1$}&\multirow{4}{*}{$V=223.5$, $p<0.01$}\\ 
&information with pen and paper & & & & &\\
\cline{2-6}
&It was very easy to select the location from recorded  &\multirow{2}{*}{$7$}&\multirow{2}{*}{$6$}&\multirow{2}{*}{$5.4$}&\multirow{2}{*}{$2.5$}& \\ 
&information with the tablet  & & & & &\\
\cline{2-7}
&I think that it would be easier for professional hackers  &\multirow{2}{*}{$10$}&\multirow{2}{*}{$9.5$}&\multirow{2}{*}{$8.5$}&\multirow{2}{*}{$2.1$}&\multirow{2}{*}{-}\\
&to gain geopass through shoulder surfing & & & & &\\
\bottomrule
\end{tabular}
\label{tab:ss_feedback}
\vskip -0.33cm
\end{table*}
 
In our study, each participant who failed to log in successfully made $10$ attempts, which was also the maximum number of attempts allowed. For each unsuccessful participant, we measured the distances (in kilometers) between the actual geopass and her selected locations. Then we selected the closest of these 10 attempts to the actual geopass and recorded this minimum distance (\textit{min-distance}).

Figure~\ref{fig:ss_distance} shows an empirical CDF (ECDF) of min-distances, where we find that the min-distance was no greater than $0.05$ kilometers for $50$\% of participants in Pan-P\&P, $40$\% in Pan-Tab, $80$\% in Typ-P\&P, and $10$\% in Typ-Tab. Overall, $48$\% of participants had a min-distance of no greater than $0.05$ kilometers. Note that here we consider those participants only who failed to log in successfully after the maximum number of allowed attempts. 

We do not get matched pairs of subjects while evaluating the difference in min-distances for failed logins of two study conditions, so we conducted a Wilcoxon-Mann-Whitney test. We did not find any significant difference between any pair of conditions for min-distances. 


\subsection{User Opinion and Feedback}\label{shoulder_feedback}

We asked the participants to answer a set of ten-point Likert-scale questions ($1$ indicating \textit{strong disagreement} and $10$ indicating \textit{strong agreement} with the given statement) about the ease of recording GeoPass through shoulder surfing and logging in using the recorded information. The results are shown in Table~\ref{tab:ss_feedback}. 

The majority of participants strongly agreed that it would be easy for a professional hacker to gain a geopass through shoulder surfing (mode 10, median 9.5). They generally found it easy to record the geopass with pen and paper (mode 8, median 8.5) and then select the location as recorded with pen and paper (mode 10, median 8).

Most of the participants found shoulder surfing with pen and paper easier than that with a tablet. We conducted Wilcoxon signed-rank tests (appropriate for matched pairs of subjects) to evaluate the differences between pen and paper and tablet in terms of user feedback on the ease of shoulder surfing using the corresponding material. We found significant differences for both ease of recording ($V=304.5$, $p<0.01$) and logging in ($V=223.5$, $p<0.01$). Participants offered a clear rationale for the distinction: In response to an open-ended question about their experience, several participants indicated they could use pen and paper to take notes while looking at the experimenter's screen, while using the tablet generally required looking at the tablet screen.


\section{Discussion}\label{disc}
In this section, we summarize and discuss the findings from our three independent studies that lead us to our conclusions about the appropriate applications for GeoPass and areas for improvement in future research. 

\subsection{Summary of findings}
\textbf{Field study.} We evaluated the usability of GeoPass in a real-world setting through a 66-day-long field study including $1781$ login sessions from $50$ participants. The study shows a satisfactory memorability for GeoPass with an overall login success rate of $96.1$\% when users have to remember a single geopass. 

We analyzed the login performance distribution among users for a detailed understanding of the usability issues. Our results show that the median login time for GeoPass was $19$ seconds, and $76\%$ of participants had median login times of 30 seconds or less. Although login times are long compared to textual passwords, even users who had to enter their geopass for a real-world purpose found the login times acceptable. 

Since the prior field studies on graphical passwords did not present a detailed analysis of training effect, it remains a particular interest of research community to learn how the login performances change over login sessions in a long-term field study. Our close examination on training effect found an overall improvement in login performance with more login sessions, with some modest fluctuations. For example, the median login time for GeoPass was $41$ seconds in the first login session, which dropped to $24$ seconds by the $13^{th}$ login session and then decreased to $17$ seconds by the $31^{st}$ login session. 

Chiasson et al.~\cite{passpoint3} argued that the login success rate in a lab study would be higher than that in a field study since the participants in a lab setting get primarily focused on their login attempts, but in a field study the authentication is usually a secondary task. Indeed, we found that the login success rate was $84\%$ in the first login session, which is less than the study by Thorpe et al.~\cite{geopass}\footnote{We note that direct comparison between different studies should be taken with caution.} The training effect, however, compensated for this, leading to a $94\%$ success rate at the $7^{th}$ login session and a $100\%$ success rate at the $13^{th}$ login session.

Lockout rules~\cite{strongpw} are implemented for many online accounts to ensure high security. To implement a lockout rule that is both secure and convenient for legitimate users, it is important to figure out the number of attempts an actual user would usually require to log in successfully. Our field study gives insight to this issue, as we found that $90$\% of participants made at most two attempts on average to authenticate successfully, and $100$\% of participants logged in successfully within five attempts on average. Thus, GeoPass is amenable to reasonable lockout rules.

\textbf{Interference study.} While the short-term lab study~\cite{geopass} and our field study found satisfactory memorability for a single geopass, we conducted an interference study to evaluate the login performance for multiple geopasses. We found that using multiple geopasses led to a significant interference effect that hurt memorability, where the overall login success rates were $58$\% in {\em login 1} (one week after registration) and $67$\% in {\em login 2} (two weeks after registration). 

A comprehensive survey~\cite{survey} on $25$ graphical password schemes reported that only three of these schemes have been evaluated through the interference study: Passfaces~\cite{interference3}, Passpoints~\cite{passpoint3,interference4}, and VIP~\cite{VIP}. The studies~\cite{interference3,passpoint3,interference4,VIP} showed the memorability for multiple graphical passwords, and reported the interference effect through comparing to a single password~\cite{passpoint3} or presenting different distracting conditions~\cite{interference3}. None of these studies~\cite{interference3,passpoint3,interference4,VIP} demonstrated, however, the precise impact of interference. In particular, they did not distinguish the login attempts that failed because of confusion with other passwords (i.e. likely interference) from attempts that failed due to simply forgetting the desired password (i.e., non-interference).

In our study, we provide a clear distinction between interference and non-interference, which reveals that $44.8$\% of attempts in {\em login 1} and $38.2$\% of attempts in {\em login 2} failed because of interference effects. We investigated both accurate and non-accurate interferences for an in-depth analysis of interference effects. As noted by Biddle et al.~\cite{survey}, how to best evaluate multiple password interference still remains an open issue; our methodology for analyzing interference effect should make an important contribution in this regard.

Our investigation on the cause of interference suggests that the participants did not seem confused by geopasses that were geographically close, since the interference effect between a pair of geopasses had no correlation with the distance between them. It remains unclear why interference occurs. Perhaps with more experience, users would learn how to associate the account with the geopass, much as they associate accounts with textual passwords today. Studies are needed to see if this is the case, for GeoPass as well as graphical and other schemes.

While prior studies~\cite{pw_adv13} showed the effectiveness of printed password advice in a lab setting, it would be of significant interest to see how the participants react to verbal password advice from security researchers while they create an authentication secret. We examined the issue in both real-life scenario (field study) and lab setting (interference study). At the beginning of studies, we verbally gave a simple security suggestion to the participants that choosing their home, school, or workplace as geopasses may make their credentials vulnerable to targeted guessing attacks. However, we found that $44.9$\% of geopasses in field study and $44.5$\% of location-passwords in interference study represent their home, school or workplaces. Our findings give a primary indication that the verbal security suggestions, even from security personnel, may not be sufficient to restrain users from creating predictable authentication secrets. Further studies are required to investigate this issue in greater detail.

\textbf{Shoulder-surfing study.} The findings from our shoulder-surfing study indicate that GeoPass is not resilient against this attack, with a $48\%$ overall success rate for participants playing the role of attackers. We closely examined the attempts of participants who failed to log in successfully to understand their possibilities in an alternate scenario. Our analysis suggests that if the unsuccessful participants were allowed more login attempts the success rate could further increase, since their clicked locations were very close to the actual geopass in several cases. As shown by our results, overall $48$\% of participants who failed to log in successfully had their best attempt within $0.05$ kilometers of the actual geopass. We also note that a resourceful and experienced attacker could be more successful in such cases.

Using the Google map interface on a tablet to mimic the navigation of a user to her location-password represents a tech-savvy method for recording geopass, while pen and paper is traditionally used in shoulder surfing studies to record credentials. Our study showed that the participants preferred pen and paper to using a tablet as they could easily take notes with pen and paper while looking at the experimenter's screen. 

For an authentication scheme, it is of particular interest to the research community to see how the user behavior during login affects the risk of observation attacks (e.g., shoulder surfing). To explore deeper into the issue, we analyzed the impact of navigation strategies on shoulder surfing, and we found that how a user navigated to the location-password, either by panning or typing the full address, did not affect the success rate of the attack. 

The success rate of university students in our shoulder-surfing study raises a red flag for not only geographical location-password schemes, but also for any graphical password schemes with mouse input~\cite{passpoint1,ccp,pccp2,passface,story}, and shoulder-surfing studies should be conducted on them before deployment.

\subsection{Application of GeoPass}\label{app}
In light of our findings, we believe that GeoPass is not yet mature enough for broad commercial deployment and demands careful attention from the research community to offer higher memorability for multiple location-passwords and better resilience against shoulder-surfing attacks. 

In its current state, GeoPass may be deployed in limited conditions, such as for the authentication of employees in an internal system where a user remembers a single geopass and usually authenticates in an environment with limited risk of shoulder surfing. 

Because of the dangers of shoulder surfing and the relatively long time for a login session, GeoPass may be best suited as a backup authentication mechanism for password resets. The high memorability for a single geopass may make it a good alternative to security questions, assuming that the issue of interference is addressed. Further research is needed to evaluate GeoPass for such an application based on longer-term recall.

\section{Conclusion}\label{conc}
In this paper, we presented a systematic approach to perform a comprehensive study on an authentication scheme to reveal the usability and security issues that need to be addressed before commercial deployment. We applied this approach to study GeoPass, a scheme that had been shown to have good usability properties and resilience to guessing. Our results show that GeoPass, while offering good real-world usability, suffers from interference issues and vulnerability to shoulder surfing. We have denoted the practical applications for GeoPass in its current state and identified the issues that need to be addressed in future research for wide-scale deployment of digital-map-based authentication schemes.






\section{Acknowledgement}
This material is based upon work supported by the National Science Foundation under Grant No. CNS-1117866 and CAREER Grant No. 0954133.

\bibliographystyle{IEEEtran}      
\bibliography{IEEEabrv,refs}
\balance

\end{document}